**Title: Motor skill learning by increasing the movement planning horizon**

**Authors**

Luke Bashford[1,2*], Dmitry Kobak[1,2,3*], Carsten Mehring[1,2]

[1]Imperial College London, London, UK

[2]Bernstein Center Freiburg, Alberts-Ludwigs University, Freiburg, Germany

[3]Champalimaud Centre for the Unknown, Lisbon, Portugal

*Equal contribution

**Corresponding author**

Luke Bashford, luke.bashford11@imperial.ac.uk

# Abstract

We investigated motor skill learning using a path tracking task, where human subjects had to track various curved paths as fast as possible, in the absence of any external perturbations. Subjects became better with practice, producing faster and smoother movements even when tracking novel untrained paths. Using a "searchlight" paradigm, where only a short segment of the path ahead of the cursor was shown, we found that subjects with a higher tracking skill took a longer chunk of the future path into account when computing the control policy for the upcoming movement segment. We observed the same effects in a second experiment where tracking speed was fixed and subjects were practicing to increase their accuracy. These findings demonstrate that human subjects increase their planning horizon when acquiring a motor skill.

# Introduction

Acquisition of motor skill is a difficult and controversial topic that remains poorly understood. Recent research on motor learning has been mostly focused on motor adaptation following a visuomotor or a force perturbation [for a recent review see (Shadmehr et al. 2010)]. Even though there is no universally accepted definition of motor skill, it is clear that motor skill acquisition is very different from motor adaptation (Shmuelof et al. 2012): whereas adaptation is fast (timescale of minutes) and brings the performance back towards the baseline, skill acquisition is slow (timescale of days, weeks or years) and improves the performance as compared to the baseline. Furthermore, most of the motor skills that people acquire, at least in the adult life, are not associated with any external perturbations, but can still be notoriously difficult and take years of practice (think of learning a new dancing movement or a new acrobatic technique). Motor adaptation is now often understood in terms of forward model learning (Wolpert et al. 1995), but what computational mechanisms lie behind skill learning remains unclear.

Informally, motor skill is usually understood as a capability to perform faster and at the same time more accurate movements than other, unskilled, individuals. A task commonly used in the experiments on motor skill learning is sequential finger tapping, where subjects are asked to repeat a certain tapping sequence as fast and as accurately as possible (Karni et al. 1995; Karni et al. 1998; Petersen et al. 1998; Walker et al. 2002). Improvement in such a task can continue over days, but learning is mostly constrained to the particular trained sequence(s); in the early experiments with a single trained sequence (Karni et al. 1995) no generalization to untrained sequences was observed, and even though recent experiments with several trained sequences (Wiestler et al. 2014) do show noticeable generalization, substantial parts of the learned skill remains sequence-specific. In contrast, motor skills in real life usually encompass a very large range of different movements and movement sequences.

In this study we aimed at creating an experimental condition which would allow us to study motor learning in the absence of external perturbations and also without any repeated movement sequences. For this, we developed a path tracking task, where subjects had to track various curved paths as fast as possible. It has recently been shown that when repeatedly tracking two fixed semi-circular paths, subjects become faster and more accurate over the course of several days (Shmuelof et al. 2012). However, this increase in the speed and accuracy does not generalize to untrained paths (Shmuelof et al. 2014). Here, instead of using a fixed path, we used different paths throughout the experimental session, each having a complicated curved form. We asked the following two questions: (i) Is it possible to acquire a general skill of path tracking, i.e. to become faster at tracking any new path while maintaining accuracy? (ii) If so, then what changes in the motor system to allow this speed increase?

In particular, we ask if the speed increase can be understood in the framework of receding horizon control (Mattingley et al. 2011), which is a restricted version of optimal control. Human motor behavior can often be well described by optimal control models (Todorov and Jordan 2002, Braun et al. 2009, Diedrichsen et al. 2010), but they may require demanding computations that the human motor system might not be capable of. In order to preserve the benefits of optimal control but reduce the computational complexity, receding horizon control, otherwise known as model predictive control, computes a feedback control policy that is optimal only for a finite planning horizon (Mattingley et al. 2011). The control policy is then continuously updated as the movement goes on and the planning horizon is being shifted forward. This allows for adaptability, e.g. if something unexpected happens during the movement. Furthermore, non-linearities may be linearized over a short horizon, simplifying optimal control computations. A recent study (Dimitriou et al. 2013) demonstrated rapid changes in feedback gains, consistent with the predictions of the receding horizon control. Our paradigm allowed us to directly measure the planning horizon of the human motor system and investigate its adaptation during skill learning.

# Results

**Experiment 1: Learning the tracking skill**

Subjects had to track various smoothly curved paths of constant length (Figure 1A) with a cursor, in the absence of any visuomotor transformations or force perturbations. They were instructed to track each path as fast as possible without touching the sides of the path; if the cursor did touch the side of the path, the trial was restarted (see Methods).

Figure 2 shows the average tracking time of successful attempts throughout the experiment for each group of subjects. Subjects were divided in three groups depending on the amount of training: naive group (red, 50 trials), learning group (blue, 300 trials) and expert group (green, 1300 trials or more). The learning effect is obvious: in each group tracking times decrease gradually across trials and form clear learning curves. For the expert group, the learning curve approached its asymptote (Figure 2A) and was almost flat on the final day of recordings (Figure 2B). There was no difference in the initial performance between the groups (mean tracking time in trials 1-5 was compared between groups, p=0.8, one-way Kruskal-Wallis ANOVA, N=21; across all subjects it was 6.6±1.6 s, mean±SD).

During training the paths were repeated in shuffled blocks of 50, but the last 50 trials used novel paths that were never presented before ("probing phase"). Still, the three groups showed markedly different performance, with the average successful time being 5.2±1.2 s (mean±SD over subjects) for the naïve group, 3.9±0.7 s for the learning group, and 2.8±0.3 s for the expert group. This difference between groups was statistically significant (p=0.001, Kruskal-Wallis ANOVA, N=21). The higher performance of the learning and the expert groups during the probing phase cannot therefore be explained by subjects memorizing specific movements and demonstrates genuine learning of path tracking.

As subjects had to restart the same trial if they hit the path side, there could be a trade-off between number of attempts and successful tracking time; riskier subjects could therefore show better average tracking times. We found that the average number of attempts differed between subjects, ranging in the probing phase from 1.2 to 2.0, but did not differ systematically between groups (Figure 2D, p=0.26, Kruskal-Wallis ANOVA, N=21) and there was no statistically significant increase with training (p=0.14, one-sided Wilcoxon-Mann-Whitney test between naïve and expert groups, N=7+7). Inspection of the relationship between mean speed and mean number of attempts across subjects (Figure 2E) suggests that there was a speed-accuracy trade-off in each group (even though correlation was not significant for any of the three groups, p>0.27) and across groups the speed-accuracy curve was shifted towards higher speeds (as confirmed by ANCOVA reporting significant group effect, p=0.0017; ANCOVA was performed on mean tracking times with two covariates: mean number of attempts and the interaction between number of attempts and the group identity, N=21). This clearly demonstrates that the improvement in tracking times with practice was not due to exploiting the speed-accuracy trade-off (i.e. higher speeds were not obtained at the cost of an increased number of attempts).

In addition to tracking time, we looked at trajectory smoothness as another measure of tracking skill (Figure 2F-G). Smoothness was computed as an entirely geometric measure with all timing information stripped away (see Methods). It is therefore complementary to the tracking speed. Higher values of trajectory smoothness mean smoother trajectories, with 1 corresponding to a straight line. As the learning proceeded, subjects in all groups produced increasingly smooth trajectories (Figure 2F-G), with significantly different smoothness in the probing phase across groups (p=0.0007, Kruskal-Wallis ANOVA, N=21), Figure 2H.

**Experiment 1: Searchlight probing**

To unravel the mechanisms of skill acquisition we designed two sorts of probing trials, which were intermixed in the probing phase (see Methods for details). The first type of probing trials were "searchlight trials", during which subjects had to track curved paths as usual, but could only see a certain part of the path (fixed distance L) ahead of the cursor, see Figure 1B. As they were moving the cursor along the path, the path was gradually appearing on the screen. The searchlight length L varied between 10% and 90% of the whole path length (i.e. the minimal L was 3.1 cm) to probe subjects planning horizon.

The results of the searchlight test are shown on Figure 3A. For each trial we calculated the mean tracking speed as the path length divided by the tracking time, and then averaged the speeds exhibited by each group of subjects at each searchlight length L. As expected, subjects in all groups moved faster with longer searchlight lengths and for all searchlight lengths the expert groups showed the highest speed, followed by the trained and then the naïve groups.

Inspection of Figure 3A suggests that expert subjects were strongly handicapped by short searchlights (they moved much slower at L=10% than at L=100%), whereas naïve subjects were moving almost as fast with short searchlights as they did when the full path was visible. Indeed, whereas in all groups tracking speed at L=10% was lower than at L=20% (p=0.008 in each case, one-sided paired Wilcoxon signed rank test, N=7+7), only in the expert group the speed further increased between L=20% and L=30% (p=0.016), and between L=30% and L=40% (p=0.008).

To quantify this effect on a single subject level, we defined the "planning horizon" for each subject as the maximal searchlight length $L_{max}$ such that tracking speeds at all searchlight lengths up to and including $L_{max}$ are significantly lower than those at L=100% at p<0.05 level (one-sided Wilcoxon-Mann-Whitney ranksum test, N=10+50 for each comparison). Accordingly, planning horizon shows the largest horizon length at which the performance is still distinguishable from the asymptotic performance. Further, for each subject, we used mean tracking speed in the probing phase (at L=100%) as a proxy for their acquired skill; the higher the tracking speed, the higher the skill. The relation between the planning horizon and the tracking skill is shown in Figure 3B. There was a strong positive correlation, R=0.65 (p=0.001, N=21, Spearman rank correlation is used here and below), and also a significant difference in planning horizons across groups (p=0.005, Kruskal-Wallis ANOVA, N=21) and significant improvement in the expert group as compared to the naïve group in particular (p=0.003, one-sided Wilcoxon-Mann-Whitney ranksum test, N=7+7). Together, this demonstrates that the more skilled subjects have larger planning horizons.

As a consistency check we used an alternative procedure that did not rely on formal hypothesis testing to find the planning horizon for each subject. Namely, we used the maximal searchlight length $L_{max}$ such that mean tracking speed at all searchlight lengths up to and including $L_{max}$ was lower than 90% of the mean speed at L=100%. This led to the same conclusions as above (correlation between planning horizon and tracking speed was R=0.52, p=0.02, N=21; Kruskal-Wallis ANOVA yielded p=0.049 for difference across groups and one-sided Mann-Whitney-Wilcoxon test between the naïve and the expert groups yielded p=0.02). The outcome did not change for alternative cutoffs such as e.g. 80% (p<0.0001, p=0.001, p=0.0003, same tests).

We stress that this effect cannot be trivially explained by the difference in speeds. Hypothetically it may have been possible that all subjects reached their asymptote performance at e.g. L=20% searchlight. This would have made the planning horizon identical for all subjects, despite very different tracking speeds. In reality, we saw a strong correlation between tracking speed and planning horizon. See also Experiment 2 below, where subjects' speed was held fixed.

In addition, we defined the "searchlight sensitivity" as the ratio of the mean tracking speed at L=100% and at L=10%, and computed it for each subject separately, see Figure 3C. Here again we found a strong and significant correlation with the mean tracking speed (R=0.78, p<0.0001, N=21), significant difference across groups (p=0.001) and significant improvement in the expert group as compared to the naïve group (p=0.0003). Together, this shows that more skilled subjects had a higher searchlight sensitivity, i.e. were more strongly handicapped by short searchlights.

In contrast, trajectory smoothness (Figure 3D-F) did not seem to depend that much on the searchlight length, note in particular nearly flat lines on Figure 3D.

We also investigated the dependency of tracking speed on the "searchlight time" (time T needed to cover the searchlight length L) instead of the searchlight length. We defined planning "time horizon" in full analogy with the definition of the planning "distance horizon" above (T was computed as L/V for each subject using their average tracking speed V in each condition). Figure 4 shows that the dependence of this measure on subject's skill is weak; still, median time horizon was smaller for the naïve group than for the expert group (p=0.049, one-sided Wilcoxon-Mann-Whitney ranksum test; median time horizons 0.6 s and 1.0 s for the naïve and the expert groups). We devised Experiment 2 to address this issue in more detail.

**Experiment 2: Searchlight probing**

In Experiment 1 varying tracking speeds of different groups led to little difference in the planning horizon in terms of time, even though it was very different in terms of distance. Therefore we designed a second experiment where subjects had to a track a path moving towards them at a fixed speed (Figure 1E and Methods). In contrast to Experiment 1, here the cursor was allowed to exit the path, but the subjects were instructed to try to stay inside the path for as much time as they could. Our main performance measure is the fraction of time that the cursor spent inside the path boundaries.

One group of subjects (the expert group) was trained for 30 minutes on each of 5 consecutive days. Another group (the naïve group) did not have any training at all. Both groups then performed a probing block of 30 one-minute-long trials with searchlights varying from L=10% to L=100% (three repetitions of each value of L). The average accuracy at full searchlight L=100% was 80.2±6.6% for the expert group and 53.9±7.5% for the naïve group (mean±SD across subjects), with the difference being highly significant (p=0.0002, Wilcoxon-Mann-Whitney ranksum test), see Figure 5A. This demonstrates that the expert group acquired the tracking skill.

We analyzed the results of the searchlight probing in exactly the same way as above. For each subject, we found the planning horizon as the maximal searchlight length $L_{max}$ such that accuracies at all searchlight lengths up to and including $L_{max}$ are significantly lower at p<0.05 level than those at L=80%, 90%, and 100% pooled together (one-sided Wilcoxon-Mann-Whitney ranksum test, N=3+9 for each comparison). Pooling was necessary as otherwise there was not enough data points for the ranksum test, but as can be seen on Figure 5A, all subjects seem to have reached the asymptote performance by L=80%. For consistency, we used the median accuracy at L=80%-100% (median across N=9 values) as a proxy for the acquired skill.

The relation between the planning horizon and the tracking skill is shown in Figure 5B. There was a strong positive correlation between planning horizon and tracking skill, R=0.78 (p<0.0001, N=20), and also a significant difference in planning horizons between groups (p=0.001, Wilcoxon-Mann-Whitney ranksum test, N=10+10). The median planning horizon in the naïve group was 6 cm and in the expert group 12 cm, corresponding to the time horizons of 0.2 s and 0.4 s.

Here again, we used several alternative approaches as consistency checks. First, we split each one-minute long trial into 60 one-second-long chunks, computed the accuracy for each chunk, and then used these 60 data points per searchlight per subject to compute the planning horizon as before but with the difference that only L=100% was used as asymptote as no pooling is necessary anymore. This yielded R=0.80 with p<0.0001. Second, we defined planning horizon via the 90% threshold from the asymptote performance (see experiment 1 above); this yielded R=0.55 with p=0.01 (without splitting trials anymore). Third, for each subject we ran linear regression of accuracy on the searchlight length for L from 40% to 100% (L=40% was the median planning horizon length across all subjects). As shown on Figure 5C, the resulting slopes are strongly correlated with tracking skill, R=0.76 (p=0.0001) and are different between groups (p=0.0008). Furthermore, the expert group had significantly positive slopes, whereas the slopes for the naïve group were not significantly different from zero (p=0.002 and p=0.77 respectively, Wilcoxon signed rank test), confirming that on average the planning horizon for the expert group exceeded L=30%, whereas the planning horizon of the naïve group did not.

In summary, all the results obtained in Experiment 2 corroborate our findings from the Experiment 1. We do not report searchlight sensitivity here because this measure is not as informative in this experiment as all subjects had similar performance at the L=10% searchlight (meaning that sensitivity is almost perfectly correlated with performance for trivial reasons).

**Experiment 2: Trajectory analysis**

In Experiment 2 naïve subjects performed worse than the expert subjects at long searchlights but all subjects performed equally badly at short searchlights. What kinematic features can these differences be attributed to?

For each subject and for each probing trial, we computed the time lag between cursor trajectory and path midline (the lag maximizing cross-correlation between them). As Figures 5A&B show, the lag was ~180 ms at L=10% for all subjects and dropped to 0 ms at L=50% for all expert subjects and for 5 out of 10 naïve subjects. The other 5 naïve subjects that showed non-zero lag at large searchlights were exactly the 5 subjects with the worst performance. Negative correlation between the asymptote lag (median across L=80-100%) and the asymptote performance was therefore very pronounced (Figure 6B, R=-0.76, p=0.0001).

Next, for each path we found all segments exhibiting similar sharp leftward or rightward bends (our inclusion criteria yielded 10±4 segments per path, mean±SD). For each searchlight length L and for each subject, we computed the average cursor trajectory over all segments (N=30±6 segments per searchlight) after aligning all segments on the bend position (Figure 6C, leftward bends were flipped to align them with the rightward bends). At L=10% all subjects from both groups follow very similar lagged trajectories, resulting in low accuracy. As searchlight increases, expert subjects reach zero lag and choose more and more similar trajectories, whereas naïve subjects demonstrate a wide variety of trajectories with some of them failing to reach zero lag and some failing to keep the average trajectory inside the path boundaries. For each subject and each L, we found the turning point of the average trajectory (marked with a dot on Figure 6C) and assessed variability of the cursor position across the path movement direction at this moment (median absolute deviation across corresponding path segments, Figure 6D). This measure decreased with growing L, but for the expert group the decrease was noticeable even between L=50% and 100% (significant negative regression slope of group mean onto L, p=0.003) whereas for the naïve group it was not (p=0.8). The asymptote variability (median across L=80-100%) was negatively correlated with asymptote performance (Figure 6E, R=-0.64, p=0.003).

In summary, at short searchlights all subjects performed poorly because their trajectories were lagging behind the path. At longer searchlights the expert subjects were able to plan their movement to accommodate the bends (the longer the searchlight the better), but naïve subjects failed to do so in various respects: either still lagging behind, or not being able to plan a good average trajectory, or exhibiting a lot of variability.

**Experiment 1: Cursor jump probing**

The second type of probing trials in Experiment 1 were "cursor jump trials" (Figure 1C, see Methods for details), where we probed the visuomotor reflex in response to the cursor suddenly moving very close to the path border. Each subject experienced ~100 trials with straight horizontal paths, and the movement in half of them was error-clamped to the path middle line, i.e. constrained to a "channel". Unbeknownst and hardly noticeable to the subject, in one half of these trials in the middle of the path the cursor briefly jumped upwards or downwards, almost reaching the side of the path, i.e. close to causing a trial failure.

The corrective force that subjects exerted on the channel wall after the cursor jump onset is shown in Figure 7A for the two jump directions separately (average over subjects). In response to the cursor jump subjects exerted a force in the opposite direction, starting at ~235 ms after the jump onset (estimated as the first time point when the average force profiles for two jump directions become significantly different from each other, paired Wilcoxon signed-rank test, $p<0.05$, $N=21$) and peaking at ~419 ms (data additionally averaged over jump directions, after flipping the sign of the responses following upward cursor jumps). The time course of this force was not different between groups ($p>0.05$ for every time point between 0 and 500 ms, Kruskal-Wallis ANOVA, data pooled over jump directions as described above) and its magnitude varied strongly between subjects in a way that was not reflecting the difference in skill. In particular, for each subject we computed the average force response at 300, 350, and 400 ms and did not find any correlation with tracking speed ($p>0.28$ in all three cases), see Figure 7B.

As can be seen on Figure 7B, three subjects exhibited very large forces (average values 4.4 N, 2.7 N, and 2.5 N at 400 ms after jump onset), whereas for other subjects the average force at 400 ms was only 1.0±0.4 N (mean±SD across subjects). These three subjects also exhibited by far the largest variability in forces (standard deviations 3.5 N, 4.9 N and 7.5 N, whereas for other subjects it was only 1.1±0.4 N; median absolute deviations were high as well showing that this was not due to some outlier trials). If we exclude these three subjects from this analysis, the remaining subjects still show no correlation with tracking speed ($p>0.19$ for all three cases of 300, 350, and 400 ms after jump onset).

Additionally, with 6 subjects (2 in each group) we performed baseline measurements of the force response, measuring it with the identical "cursor jump" paradigm in the very beginning of the experiment before any path tracking. The difference between the baseline force response and the force response after skill learning is shown on Figure 7C for these 6 subjects. Across subjects there was no significant difference between the baseline force response and the force response after skill learning (p=0.4, paired Wilcoxon signed-rank test, N=6).

# Discussion

We suggested a novel paradigm that allowed us to study human motor skill learning in the absence of external perturbations: subjects had to track various curved paths either as fast as possible with the fixed accuracy (Experiment 1) or as accurately as possible with the fixed speed (Experiment 2). In Experiment 1 the cursor was not allowed to cross path borders and as the number of trial attempts did not increase with practice, the accuracy was essentially held fixed and so we used tracking speed as the single measure of performance (Shmuelof et al. 2012; Reis et al. 2009). We found that subjects become better (i.e., faster) with practice: only 30 minutes of practice brings substantial improvement and after 5 days of training performance approaches its asymptote (Figure 2). In contrast, in Experiment 2 the tracking speed was fixed and so we used the accuracy, i.e. the fraction of time the cursor was inside the path boundaries, as the measure of performance. Here again we observed substantial improvement after 5 days of training (Figure 5). At the same time, in either experiment subjects' movements were not perturbed in any way (e.g. no visuomotor transformations and no force fields were applied), so the improvement in performance cannot be due to adapting an internal model to compensate for an external perturbation (Shadmehr et al. 2010). In addition, the paths were different on every trial, so the improvement in performance cannot be attributed to motor memory either.

**Planning horizon increases with motor skill**

We ask therefore what changes in the motor system occur during the time scale in which we observed learning that allowed skilled subjects to perform better? We hypothesized that a large role in our tracking tasks is played by taking into account approaching path bends and preparing for an upcoming movement segment. Skilled subjects could become more accurate by increasing the amount of future path that is taken into account and by using the available information more efficiently. Note that "preparing" for the movement can be interpreted differently depending on the computational approach. In the framework of optimal control (Todorov and Jordan, 2002) subjects do not plan a trajectory to be followed, but instead compute the optimal time-dependent feedback policy and then make the movement according to this policy.

We tested this hypothesis with a "searchlight" probing. In Experiment 1 subjects had to track various paths but only saw a short segment of the path ahead of the cursor; the path was being built up as the cursor proceeded along. We found that subjects with a higher tracking skill demonstrated (a) larger planning horizon and (b) higher searchlight sensitivity (Figure 3). A larger planning horizon means that skilled subjects' performance could be still be impaired at larger searchlight lengths, i.e. skilled subjects took a larger chunk of path into account when preparing for the upcoming movement segment, e.g. by computing a feedback policy (~10 cm for the expert group, as opposed to only ~3 cm for the naïve group). Indeed, processing a 10 cm chunk of lying ahead path is computationally more demanding than processing only a 3 cm chunk, considering that parameters of the paths like curvature etc. remained the same across searchlight lengths. Higher searchlight sensitivity means that skilled subjects were more strongly handicapped by very short searchlights than the subjects with poor tracking skills and is a complementary effect to the increased planning horizon.

Planning horizon measures the length of the path chunk that subjects use for planning the movement, or the "look ahead distance". As expert subjects also move faster, it makes sense to also assess the planning time horizon, i.e. the "look ahead time". This time horizon was also larger for the expert group (~1 s, as opposed to ~0.6 s for the naïve group), corroborating our interpretation, though the difference for the time horizon was less pronounced and only weakly significant. To demonstrate more clearly that the time horizon can increase with learning a tracking skill, we therefore devised a separate experiment where subjects had to track paths at a fixed speed. In this Experiment 2 we again found that subjects with a higher tracking skill demonstrated larger planning horizons: ~12 cm for the expert group vs. ~6 cm for the naïve group, corresponding to the time horizons of ~0.4 s and ~0.2 s.

The decrease in performance at very short searchlights in both experiments can be easily explained by various factors. For example, in Experiment 1 it might be that subjects are slowing down at short searchlights to avoid hitting a path boundary if a sharp bend comes from behind the searchlight, similar to a cautious car driver "driving in the fog". And in Experiment 2 subjects seem to have very low accuracy at short searchlights because their reaction time is not enough to follow the visible path segment rapidly moving from left to right. These effects, however, do not explain the difference between naïve and expert subjects, and the dependence of the horizon length on subject's skill acquired with practice.

The observed increase in planning horizon can be interpreted in the framework of model predictive control, also known as receding horizon control, RHC (for a review, see Mattingley et al. 2011). In RHC, the optimal control policy (Todorov and Jordan 2002) is computed for a finite and limited planning horizon, which may not capture the whole duration of the trial. This policy is then applied for the next control step, which is typically very short, and the planning horizon is then shifted one step forward to compute a new policy. Hence, RHC does not use a pre-computed policy, optimal for an infinite horizon, but a policy which is only optimal for the current planning horizon. Increasing the length of the planning horizon is therefore likely to increase the accuracy of the control policy. In our experiments this would allow for faster movements without increasing the number of attempts. RHC is a well established control framework in engineering, however, it has not yet been used in movement neuroscience. We propose that our results can be framed in the general context of RHC as we suggest that subjects with a higher skill have a longer planning horizon, indicating that subjects learn how to take advantage of future path information to improve motor performance.

This increase in the planning horizon, however, is unlikely to account for all of the observed improvement in performance: note that in Experiment 1 the trained subjects perform better than naive already at the shortest searchlight length, Figure 3A (even though this difference is very small in Experiment 2). Evidence for an additional factor is provided by of the analysis of movement smoothness. Skilled subjects produced smoother trajectories (Figure 2H) but the trajectory smoothness depended only very weakly on the searchlight length (Figure 3D). We suggest that our smoothness measure reflects the level of execution noise in the motor system. Execution noise here includes contributions ranging from motor noise in motor neurons and muscle fibers to neuronal networks in higher areas involved in computing the motor command according to the current policy. Our findings indicate that the reduced execution noise is an additional correlate of skill acquisition, which is complementary to and largely independent from the improved movement planning. Apart from the two processes discussed here – increased planning horizon and reduced execution noise – further motor and non-motor processes may play a role in tracking skill learning, which remain to be investigated in future studies.

**Visuomotor reflex reactions**

One of our a priori hypotheses was that skilled subjects learn to "recover" the trial in situations when the cursor approaches the path side. We tested this hypothesis in Experiment

Page **20** of **45**

1 by using a cursor jump paradigm, where the cursor unexpectedly jumped to the side while subjects were making movements in a force channel. It is well known that subjects produce a corrective force as a reflex reaction to a cursor jump and that the feedback gain of this reflex can be adapted during learning or according to the environment (Franklin and Wolpert 2008; Franklin et al. 2012; Kobak and Mehring 2012; Dimitriou et al. 2013). Nevertheless, in our task we found that these reactions were not different between the naive, trained and expert groups and, over subjects, the reflex gain did not correlate with the tracking skill (Figure 7B). In addition, for a subset of subjects we made baseline measurements of the force magnitude, and across subjects there was no difference in force before and after skill learning (Figure 7C).

We note, however, that the absence of changes in the feedback gains might be due to a limitation in our test procedure: we used straight paths to probe the reaction to perturbations, but curved paths to train the tracking skill (because we found it not possible to make a curved but unnoticeable force channel and so could not use curved paths for the cursor jump trials). It has been recently demonstrated that feedback gains can adapt rapidly to task changes (Dimitriou et al. 2013) and it is, therefore, possible that the gains measured with straight paths are different from the gains used while tracking curved paths. Also, in a recent study where subjects were trained to track two fixed semi-circular paths (Shmuelof et al. 2012) the authors concluded that feedback control did improve with training.

In the case of straight paths receding horizon control with different horizon lengths would arguably result in identical feedback gains. Therefore, the lack of noticeable difference in feedback gains between groups does not contradict our interpretation in terms of the receding horizon control.

**Previous studies on path tracking**

Even though our study is the first to investigate skill learning with tracking different paths, similar approaches were used before. In the 1950–1970s extensive research was carried out on tracking tasks in humans (Poulton 1974); in the usual experimental setting subjects would track a curve drawn on a paper roll moving with a fixed speed, similar to our Experiment 2. Poulton observed that the accuracy of the tracking increased with practice and also found that it increased with the searchlight length (which was modified by physically occluding part of the paper roll) (Poulton 1974, p 187). These studies, however, did not investigate the effect of learning on the planning horizon. In a more recent study subjects had to track a fixed maze without visual feedback and learnt to do it faster as the experiment progressed (Petersen et al. 1998); there the task was partially cognitive as subjects had to "discover" and then remember the correct way through the maze.

A path tracking task similar to ours was investigated by Shmuelof et al. (2012, 2014), who used two fixed semi-circular paths. A crucial difference between these studies and ours is that we used different paths throughout the experiment and investigated the generalization of the path tracking skill to novel paths. This makes our task resemble structure learning experiments, where subjects are confronted with multiple visuomotor or force perturbations, and learn the invariant relationships between perturbation parameters (Braun et al. 2009; Kobak and Mehring 2012; Yousif and Diedrichsen 2012). In the present experiment, every single path can be seen as a separate motor task, and the common control principles required to track any curved path can be hypothesized to form a structure.

**Conclusion**

In conclusion, we have established that people are able to learn the skill of path tracking and achieve seemingly final performance after 5 days of training. This increase in motor skill is associated with the increased reliance on planning future movement segments, and shows an analogy to receding horizon control.

## Materials and Methods

**Experiment 1: Subjects**

Twenty naive volunteers (16 males and 4 females, age range 19-29 years old) participated in this experiment. Subjects gave informed consent and were paid £5 for each day of the experiment. One subject was excluded from the analysis because he was obviously not paying attention during the experiment and demonstrated no learning. Additionally, two authors (LB and DK) participated in one of the subject groups (see below), without payment. The experiment received ethics approval by Imperial College London.

**Experiment 1: Experimental setup**

Subjects were seated in front of the horizontal desk and with their dominant hand held the handle of a robotic manipulandum (SensAble Phantom 3.0). To allow frictionless movements, the handle was mounted on an air sled on the horizontal glass surface. Another horizontal surface ~20 cm above was used to display the workspace that was projected from above using a standard projector. Subjects could not see their hand, but the system was calibrated such that the cursor was always displayed directly above the Phantom handle. The distance between subjects' eyes and the centre of the workspace was 40-45 cm depending on subject's height. Subjects were moving the handle by moving their whole arm.

**Experiment 1: Task**

To start each trial subjects had to move the cursor ($R$=2.5 mm) to a central target ($R$=9.4 mm) and to hold it there for 1 second. Then a curved path appeared on the screen and subjects had to track it with the cursor as fast as possible; each path was 1.88 cm wide and 31.4 cm long (Figure 1A). If during the movement the cursor touched the side of the path, the path immediately disappeared and a message "Careful!" appeared on the screen, accompanied by a short buzzer sound. The central target was displayed again, allowing the trial to be reattempted. Only when each path was successfully completed was the next one displayed. The tracking time of the last successful trial, the overall number of failed attempts, the overall tracking time and the number of remaining paths were displayed throughout the experiment. The subjects were instructed to complete the paths as fast and as accurately as possible and were encouraged always to try to beat their fastest completion time.

**Experiment 1: Paradigm**

The experiment consisted of a training phase and a probing phase (Figure 1D). During the training phase paths were presented in blocks of 50 trials, each block being a random permutation of the same 50 pre-generated paths. Subjects were divided into three groups depending on the amount of training (see Figure 2A). The naive group (N=7) experienced one block of 50 training trials. The learning group (N=7) experienced 6 blocks, i.e. 300 training trials, with a 5 minutes break after the first 5 blocks. Finally, the expert group (N=7) had the same paradigm as the learning group, but 4 subjects additionally experienced 20 training blocks (1000 training trials) over 4 preceding days (5 blocks per day) and 3 subjects (including the authors LB and DK) had comparable or even exceeding amount of training over the course of preceding several months. We did not see any differences between the authors and other subjects in the expert group (see Figures; note also that on Figure 7B two green points that stand out from the rest are not the authors).

The probing phase was the same for all groups and consisted of 50 trials displaying novel paths, 90 "searchlight" trials (displaying novel paths as well) and 102 straight trials with "cursor jumps". This entire set of 242 probing trials was split in two equal batches, the order inside each batch was randomized, and they were presented to the subjects with a 5 minutes break in between (the same paths in the same pseudorandom order were presented to all subjects).

In the "searchlight" trials subjects could only see a segment of the path up to a certain distance ("searchlight length") ahead of the cursor; as they were moving the cursor along the path, the path gradually appeared on the screen (Figure 1B). We used 9 different searchlight lengths from 10% to 90% of the full path length in steps of 10%; each length was used 10 times, making 90 trials in total. Again, 90 novel paths were used for these trials.

In the "cursor jump" trials subjects had to track a straight left-to-right path (Figure 1C). Among these trials there were 50 normal trials and 52 error-clamped trials with cursor jumps (26 jumps upwards and 26 jumps downwards). During the error-clamp trials subjects moved in a simulated "channel" with Phantom applying a restoring spring-like force towards the path middle line (spring constant k=2500 N/m). When subjects were 8 cm away from the starting position, cursor jumped 5 mm either up or down and was moving for 5.5 cm parallel to subject's hand before jumping back. The straight path had the same length as the curved paths, but a different starting point, because otherwise it would not fit into the workspace. Note that in all trials the center of the cursor was not allowed to be further than 6.9 mm from the middle line (otherwise it touched the side of the path and the trial had to be restarted); the average distance from the middle line at the time of jump onset was 0.6±0.9 mm (mean±SD across all subjects), thus the jump magnitude of 5 mm brought the cursor very close to the path edge. Larger jumps would often have brought the cursor beyond one of the path boundaries, a situation that was not allowed in our task.

Six subjects (2 in each group) experienced an additional block of "cursor jump" probing trials (exactly as described above, 102 trials) in the very beginning of the experiment, before starting to track any curved paths. They had a 5 minutes break after this introductory session. This was done to assess the subjects' baseline performance. Subjects in the expert group had this baseline probing in the beginning of day 1.

**Experiment 1: Path generation**

All paths were generated in advance by the following algorithm. Each path consisted of 1000 points $\{x_i\}$, with $x_1$ always being a fixed centre position. The distance between any two successive points $x_i$ and $x_{i+1}$ was 0.314 mm, so the length of each path was approximately S=31.4 cm. Let us designate by $\alpha_i$ the polar angle of the vector $x_{i+1} - x_i$; the sequence $\{\alpha_i\}$ was first generated as a random walk starting with a random angle $\alpha_1$ and with step size distribution being uniform between ±12°, and then smoothed with a 3rd order Savitzky-Golay filter with a window size of 201. The sequence $\{x_i\}$ was then reconstructed given $x_1$ and $\{\alpha_i\}$. Among the paths generated by this algorithm we manually selected the ones not crossing the workspace borders and without self-intersections. See Figure 1A for several exemplary paths.

**Experiment 1: Data recording and analysis**

Cursor position and produced force were recorded at 1000 Hz (occasionally missing values were filled in with linear interpolation), and low-pass filtered at 50 Hz ($3^{rd}$ order Butterworth filter). To compute force responses following cursor jumps (Figure 7), we performed trial-wise baseline correction by subtracting the baseline force in each trial (average force from 0 to 100 ms after jump onset).

For each trial, tracking time T was defined as the time between path appearance and cursor exiting the end of the path. Mean tracking speed V was defined as the path length S divided by the tracking time, V = S/T. Note that this is slightly different from the speed averaged over time, but the difference was very small because the length of actual trajectory always stayed very close to S.

We designed a measure quantifying the smoothness of the movement trajectories, independent of the dynamics of the movement (velocity, acceleration, etc.). This measure, which we termed "trajectory smoothness", was defined as follows. Let function **g**(t) be a curve that represents the movement trajectory during a trial (**g**(t) maps time $t$ to (x,y) coordinates of the cursor). The arc length of the curve between movement onset and time $t$ is given by l(t) = ∫|**g**'(z)|dz with integration from 0 to $t$. The inverse of the arc length function, i.e. $t(l)$, can be used to reparametrize the curve: **h**(l) = **g**(t(l)). This so-called natural or arc length parametrization has the property that its parameter $l$ moves along the curve at unit speed, i.e. |**h**'(l)|=1. Using this parametrization, we define an entirely geometric measure of trajectory smoothness as s = 1-c<|**h**'''(l)|>, where angular brackets represent mean value over the whole trajectory (i.e. integral with respect to $l$ over the whole trajectory divided by the arc length of the curve) and constant $c$ is set such that the least smooth movement trajectory observed across all trials and all subjects has a smoothness value of 0. A straight trajectory would have smoothness 1.

**Experiment 2: subjects**

20 naïve subjects took part in this experiment (11 males and 9 females, age range 20-34 years old). Subjects gave informed consent and were paid 10 €/h. The experiment received ethics approval from the University of Freiburg.

**Experiment 2: Setup**

Subjects sat at a desk looking at a computer monitor (Samsung Syncmaster 226BW) located ~80cm away. A cursor displayed on the screen (Matlab and Psychophysics Toolbox Version 3) was controlled by movements of a computer mouse. The mouse could be moved on the desk in all directions but only the horizontal (left and right) component of the movement contributed to moving the cursor: the vertical position of the cursor was fixed at 5.7mm above the base of the screen.

**Experiment 2: Task**

To begin each trial subjects had to press the space bar. This displayed the cursor (R=2.9mm) and the path (width = 2.83cm) going from top to bottom of the screen (30cm) and dropping continuously from the top with a vertical speed of 34.1cm/s. The initially visible path was a straight line centered in the middle of the screen with the cursor positioned in the middle of the path. Once this initial section moved through the screen, the path then followed a random curvature (Figure 1E). Subjects were instructed to keep the cursor between the path borders at all times moving only in the horizontal plane and were told to be as accurate as possible. The cursor and path were displayed in white on a black background if the cursor was within the path, and both turned red when it was outside the path.

The cursor position was sampled at 60 Hz and the tracking accuracy was defined for each trial as the percentage of time steps when the cursor was inside the path. Running accuracy values were continuously displayed in the top left corner of the screen and final accuracies were displayed in between each trial.

**Experiment 2: Paradigm**

Subjects were randomly assigned into two groups: expert and naive (N=10 in each). The paradigm included a training (expert group only) and a testing (all subjects) phase (Figure 1F). Subjects in the expert group trained over 5 consecutive days, each day completing 30 min. of path tracking (10 of 3-minute trials with short breaks in-between). If the performance improved from one trial to a next subjects saw a message saying "Congratulations! You got better! Keep it up!", otherwise the message "You were worse this time! Try to beat your score!" was shown if the performance decreased. The training paths were randomly generated on the fly. Experts performed the probing set of trials after a short break following training on the final (5th) day. Naïve subjects performed only the probing set of trials.

The probing phase lasted 30 min (30 of 1-minute trials with breaks in-between) using 30 different pre-generated paths that were the same for all subjects. The probing phase in this experiment contained 3 normal trials (L=100%) and 27 searchlight trials (L=10-90%) where some upper part of the path was not visible. Three blocks of 10 trials with the searchlight length ranging from L=10% to L=100% (in steps of 10%) were presented, with the order shuffled in each block; the same fixed pseudorandom sequence was used for all subjects.

**Experiment 2: Path generation**

Paths were generated before each trial start during training and a pre-generated fixed set was produced in the same way for testing. Each path was initialized to start at the bottom middle of the screen and the initial 30 cm of each path were following a straight vertical line. Subsequent points of the path midline had a fixed Y step of 40 pixels (1.1 cm) and an X step drawn from a uniform distribution from 1 to 80 pixels. Any step that would cause the path to go beyond the right or left screen edges was recalculated. The midline was then smoothed with a Savitzky-Golay filter (12$^{th}$ order, window size 40) and used to display path boundaries throughout the trial. All of the above parameters were determined in pilot experiments to create paths which were very hard but not impossible to complete after training.

**Statistical analysis**

In all cases we used nonparametric ranked analogues of the conventional statistical tests to avoid relying on the normality assumption. I.e. we used Spearman's correlation coefficient instead of the Pearson's one, Wilcoxon signed-rank test and Wilcoxon-Mann-Whitney ranksum test instead of one-sample and two-sample t-tests (paired and unpaired), and Kruskal-Wallis (ranked) one-way ANOVA instead of simple one-way ANOVA. We also used usual parametric tests relying on normality assumption, and all the conclusions stayed qualitatively the same.

# Acknowledgements

The study was in parts supported by the German Federal Ministry of Education and Research (BMBF) grant 01GQ0830 to BFNT Freiburg-Tübingen. The authors would like to thank an anonymous reviewer for suggesting experiment 2.

# Figures and legends

**Figure 1**

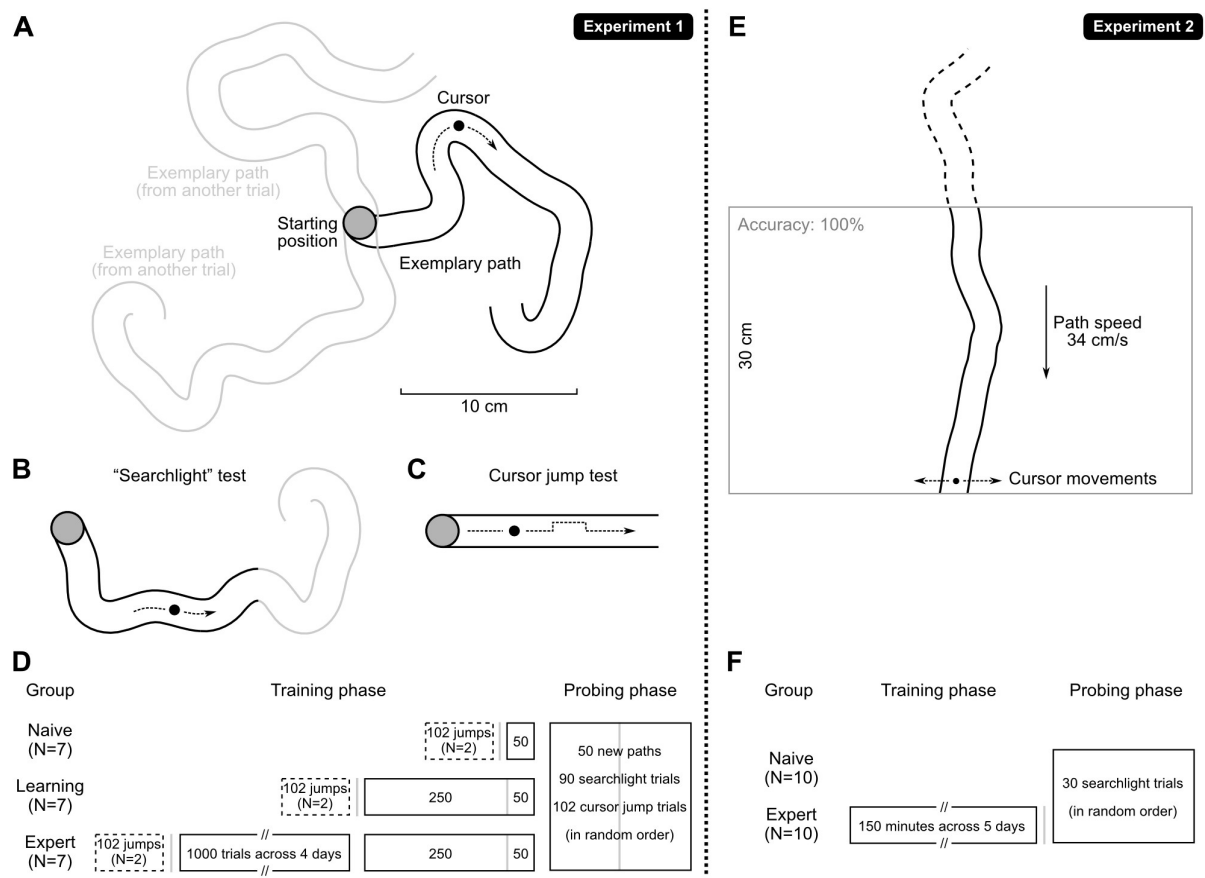

Path tracking tasks. (A) Experiment 1: Subjects had to track various curved paths with a cursor (black circle); subjects were instructed to move the cursor as fast as possible without hitting the sides. On each trial subjects saw only one path; one exemplary path is shown in black and two more exemplary paths in grey. Each path was 31.4 cm long and had the same starting location (grey circle). (B-C) Two types of probing perturbations were used: in the "searchlight" condition subjects could only see a certain length of trajectory ahead of the cursor; in the "cursor jump" condition the cursor was briefly jumping to the side, while subjects tracked a straight path in a force channel (see Methods for details). (D) Paradigm schematic for all three groups of subjects (see Methods for details). Gray vertical lines show 5 minutes breaks. (E) Experiment 2: Subjects had to track a curved path that was dropping down from top to bottom of the screen with a fixed speed of 34 cm/sec by moving the cursor horizontally. (F) Paradigm schematic for the two groups of subjects.

**Figure 2**

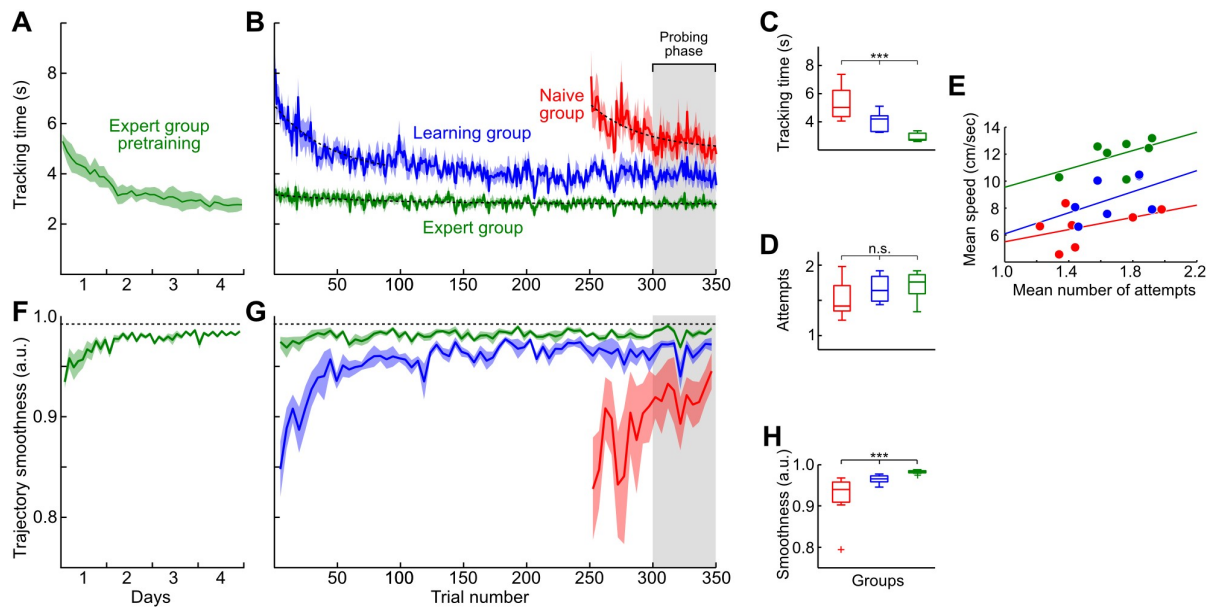

Learning curves in Experiment 1. (A) Tracking time over the first four days of training for the expert group (N=4, for other subjects data was not available, see Methods). "Tracking time" of each trial is the time of successful tracking attempt. To smooth the curve, tracking times were averaged over blocks of 25 paths. Error bars show SEM over subjects (N=4). (B) Average learning curves for each group of subjects (naive, learning and expert; for the expert group it was day 5). Shaded areas show SEM over subjects (N=7 in each group), dashed lines show exponential fits. (C) Average tracking time in the last 50 paths ("probing phase") for each group. The box plot shows median, 25th and 75th percentiles and extreme values across N=7 subjects in each group (*** p<0.001, Kruskal-Wallis ANOVA). (D) Average number of attempts in the probing phase for each group. (E) Relationship between mean number of attempts and mean tracking speed across subjects. Lines show linear fits done separately for each group. They have a positive slope in each group, indicating speed-accuracy trade-off. With training, the speed-accuracy curves are shifted upwards, demonstrating skill learning. (F) Trajectory smoothness (smoothness of cursor trajectory without taking timing into account) over the first four days of training for the expert group. Each value is the average of this measure in a block of 25 paths, shaded area shows SEM over subjects (N=4). Dashed line shows mean smoothness of the target paths (0.992±0.002, mean±SD across paths). (G) Trajectory smoothness for each group of subjects. To reduce the noise, data was averaged over blocks of 5 trials. Shaded areas show SEM over subjects (N=7 in each group). (H) Average trajectory smoothness in the probing phase for each group.

**Figure 3**

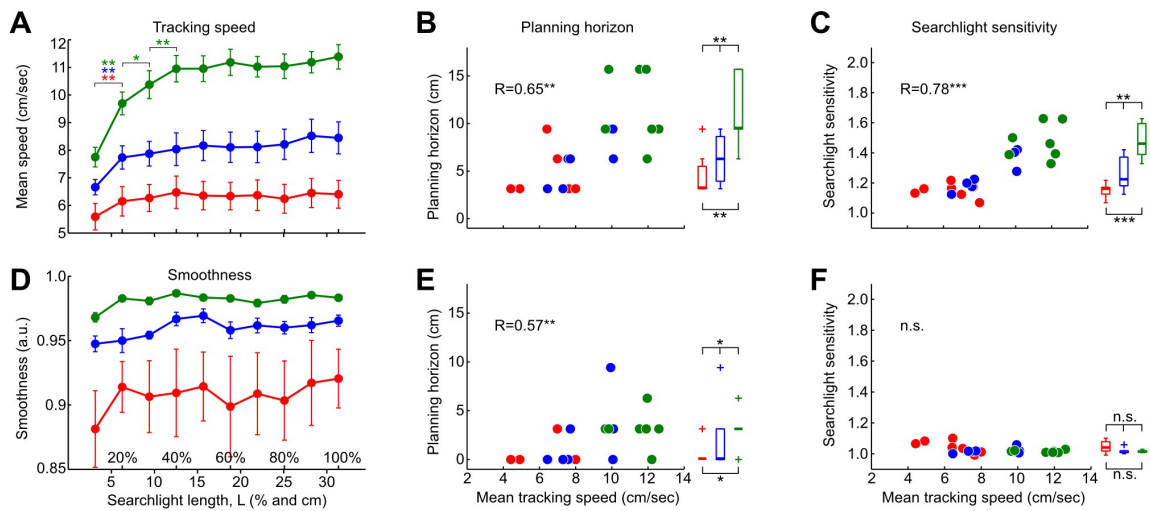

Experiment 1, searchlight trials. (A) Average tracking speeds for each searchlight length (L from 10% to 90% of the full path length) and for the full path (L=100%) for each group of subjects. Error bars show SEM over subjects (N=7). Stars mark significant differences between tracking speeds at different values of L (*p<0.05, **p<0.01, one-sided paired Wilcoxon signed-rank, N=7). (B) Planning horizon was defined for each subject as the maximal searchlight $L_{max}$ such that tracking speeds at all searchlight lengths up to and including $L_{max}$ were significantly lower that at full paths, L=100%. The scatter plot shows relation between subjects' skill (as assessed by the mean tracking speed) and their planning horizon. Correlation coefficient is shown on the plot (N=21, **p<0.01). Colour of the dot indicates the group. Bar plots on the right show group-level distributions of planning horizon. Stars on top depict the result of Kruskal-Wallis ANOVA test, stars on the bottom depict the result of one-sided Wilcoxon-Mann-Whitney ranksum test between naïve and expert groups (*p<0.05, **p<0.01, ***p<0.001). (C) Searchlight sensitivity was defined for each subject as the ratio between tracking speeds at L=100% and at L=10%. Data presentation and statistical tests same as in (B). (D--F) The same as above but for trajectory smoothness.

**Figure 4**

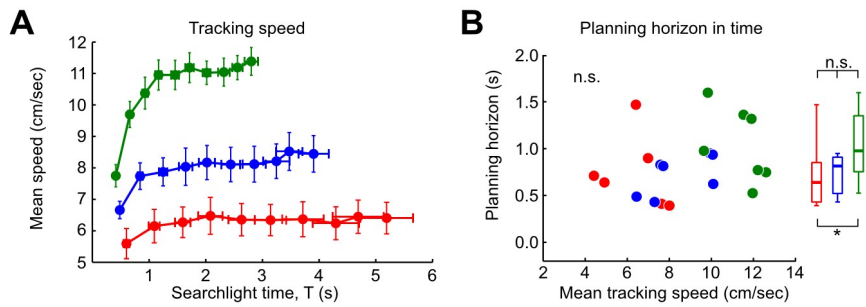

Experiment 1, planning horizon in time. (A) Whereas Figure 3A shows the dependency of tracking speed V on the searchlight length L, this figure shows the dependency of tracking speed V on the "searchlight time" T=L/V, i.e. time needed to cover the searchlight length. The vertical coordinates of each point are exactly the same as on Figure 3A. The horizontal coordinates were calculated for each subject using the average tracking speed in each searchlight condition, and then averaged over subjects Horizontal error bars shows SEM over subjects (N=7). (B) Planning time horizon, computed exactly as in Figure 3B, but using searchlight time instead of searchlight distance. Statistical analysis follows Figure 3B.

**Figure 5**

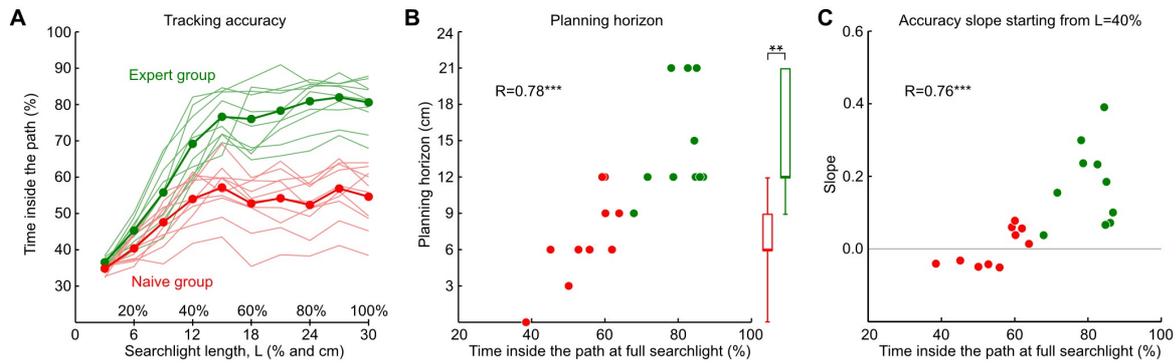

Experiment 2, searchlight trials. (A) Median tracking performance for each searchlight length for each individual subject (faint lines) and mean of per-subject values (bold lines), in red for the expert group and in green for the naïve group. (B) Planning horizon was defined for each subject as the maximal searchlight $L_{max}$ such that accuracies at all searchlight lengths up to and including $L_{max}$ were significantly lower than the asymptote performance. The scatter plot shows relation between subjects' skill (as assessed by the asymptote performance) and their planning horizon. Correlation coefficient is shown on the plot (N=20, ***p<0.001). Colour of the dot indicates the group. Bar plots on the right show group-level distributions of planning horizon. Stars on top depict the result of Wilcoxon-Mann-Whitney ranksum test between the groups (**p<0.01). (C) Accuracy slope was defined for each subject as the regression line slope of the performance curve from panel (A) from L=40% to L=100%. Data presentation and statistical tests same as in (B).

**Figure 6**

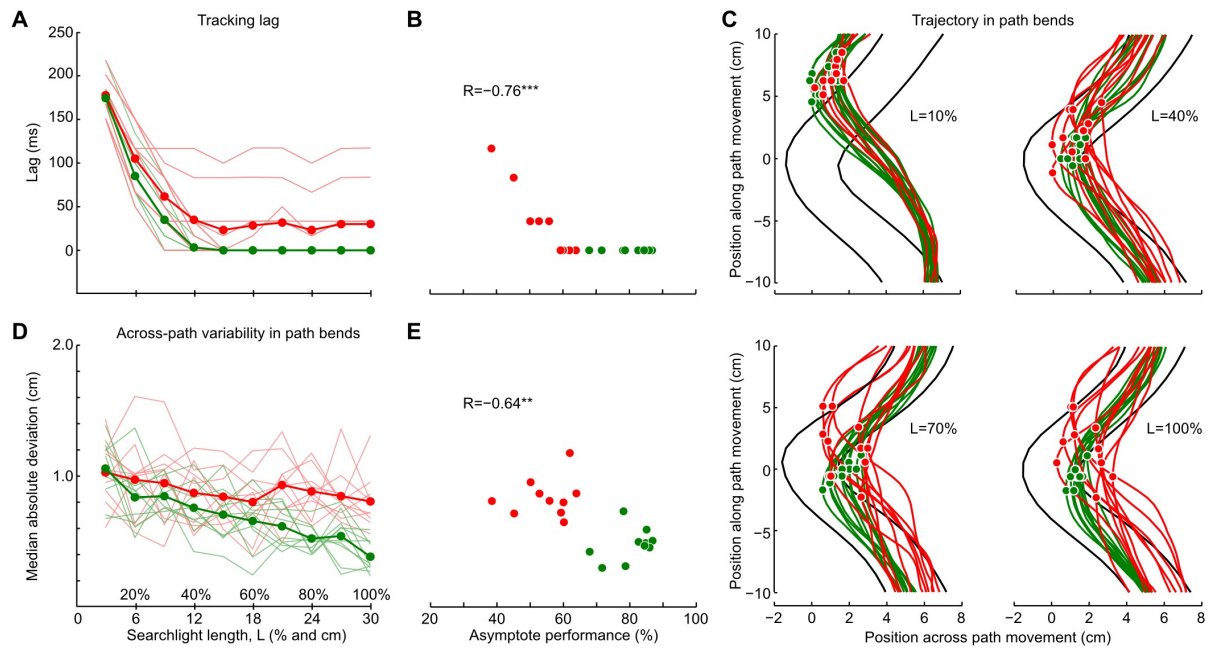

Experiment 2, analysis of trajectories. (A) Median time lag between cursor trajectory and path midline, for each searchlight length for each individual subject (faint lines) and mean of per-subject values (bold lines), in red for the expert group and in green for the naïve group. (B) Asymptote lag and asymptote performance across subjects. Correlation coefficient is shown on the plot (N=20, ***p<0.001). Colour of the dot indicates the group. (C) Average per-subject trajectories in sharp bends (leftward bends were flipped to align them with the rightward bends). Each trajectory is averaged across approximately 30 bends (the number of bends varied across searchlight lengths). Colour of the lines indicates the group. Black lines show average path contour. Dots show turning points of the trajectory. Subplots correspond to searchlight lengths L=10%, 40%, 70%, 100%. (D) Median absolute deviation of the across-path cursor position at the along-path position given by the turning point on the average trajectory in (C). Per-subject curves shown in faint lines, group means in bold lines, colour indicates the group. (E) Asymptote median absolute deviation and asymptote performance across subjects. Correlation coefficient is shown on the plot (N=20, **p<0.01). Colour of the dot indicates the group.

**Figure 7**

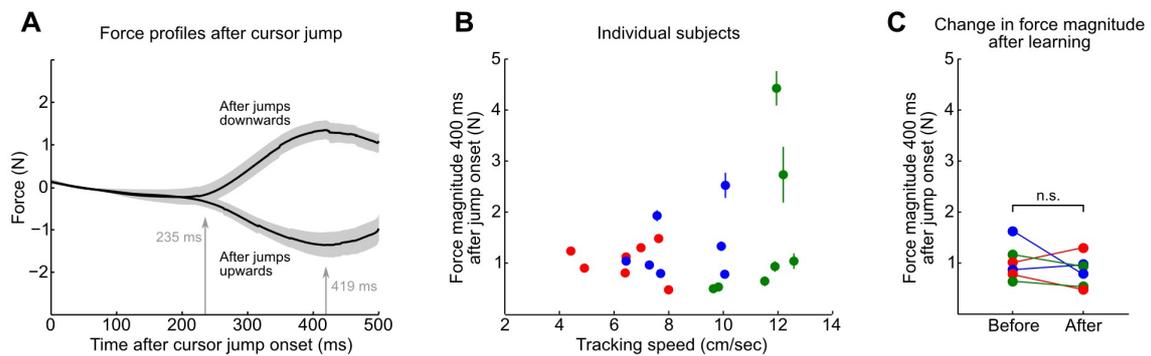

Cursor jump trials. (A) Average force profiles following a cursor jump for both jump directions, averaged over all subjects; shaded areas show SEM over subjects (N=21 for each of the two curves). The average force over the first 100 ms was subtracted from each force profile. Steps in the curves are due to some trials being faster and finishing earlier. Two arrows mark the first point when the two responses become significantly different from each other (235 ms), and the point of maximum response amplitude (419 ms). (B) For each subject we computed the value of the force response at 400 ms after jump onset (after flipping the sign of the responses following upward cursor jumps); this value is plotted here against the average tracking speed (as a proxy for acquired skill). Colour of each dot corresponds to subject's group, error bars show SEM (N=52). Most error bars are too small to be visible. (C) The difference in force response before and after skill learning is shown here for 6 subjects; across subjects, this difference is not significant.

Page 45 of 45